\documentclass[fleqn,12pt,twoside]{article}
\usepackage{espcrc1}

\usepackage{graphicx}
\usepackage[figuresright]{rotating}
\usepackage{amsmath}

\newcommand{\AmSj}{{\protect\the\textfont2
  A\kern-.1667em\lower.5ex\hbox{M}\kern-.125emS}}

\title{Finite Temperature Quantum Effects in Many-body Systems by Classical Methods}

\author{Jeffrey Wrighton\address[UF]{Department of Physics, University of Florida, \\ Gainesville, FL 32611, USA}, %
        James Dufty\addressmark,
        Sandipan Dutta\address[APCTP]{Asia-Pacific Center for Theoretical Physics,\\ Pohang 790-784, South Korea}
}       
\begin{document}

\maketitle

\tableofcontents

\begin{abstract}
A recent description of an exact map for the equilibrium structure and
thermodynamics of a quantum system onto a corresponding classical system is
summarized. Approximate implementations are constructed by pinning exact
limits (ideal gas, weak coupling), and illustrated by calculation of pair
correlations for the uniform electron gas and shell structure for harmonically
confined charges. A wide range of temperatures and densities are addressed in
each case. For the electron gas, comparisons are made to recent path integral
Monte Carlo simulations (PIMC) showing good agreement. Finally, the relevance
for orbital free density functional theory for conditions of warm, dense
matter is discussed briefly.
\end{abstract}


\section{Introduction and Motivation}

\label{sec1}A fundamental description of the thermodynamics (e.g., free
energy) and structure (e.g., pair correlation function) for materials
comprised of electrons and ions remains a challenge for many state conditions
of current interest \cite{WDM}. Typically the ions can be described by
semi-classical methods due to their relatively large masses. In contrast, the
electrons may require an accurate description of strong quantum effects.
Typical solid state conditions occur at temperatures well below the electron
Fermi temperature for which the multiplicity of zero temperature many-body
theories and simulations are available. However, as the temperature is
increased to several times the Fermi temperature such methods fail or become
increasingly difficult to implement. At still higher temperatures, 
effective classical methods can be applied (e.g., molecular dynamics
simulation \cite{MD}, Monte Carlo integration, classical density functional
theory \cite{Lutsko}, liquid state theory \cite{Hansen}). There is a long
history of phenomenological attempts to extend these classical methods to
lower temperature by including quantum effects in modified pair potentials
\cite{Murillo}. More recently, these effective classical systems have been
improved significantly by the inclusion of quantum effects in a modified
classical temperature \cite{PDW,DW,Liu14}. The entire approach of constructing
a classical system to replicate the thermodynamics and structure has been
given a formally exact context from which more controlled approximations can
be constructed \cite{DuftyDutta,DuttaDufty}. The objective here is to
summarize briefly this latter work and to illustrate its utility by two
applications: 1) the calculation of pair correlations for the uniform electron
gas, and 2) the description of shell structure for charges in a harmonic trap.
In both cases the emphasis is on conditions ranging from classical to strongly
quantum mechanical. The last section describes how this effective classical
approach can be exploited to address current problems of "warm, dense matter"
via orbital free density functional theory \cite{WDM}.

\section{Definition of the Effective Classical System}

\label{sec2}Consider a system of $N$ particles in a volume $V$ with pairwise
interactions and an external single particle potential. The Hamiltonian is
\begin{equation}
H_{N}=K+\Phi+\sum_{i}^{N}v\left(  \mathbf{q}_{i}\right)  , \label{2.1}%
\end{equation}
where $K$ and $\Phi$ are the total kinetic and potential energies,
respectively. The form of the pair potential $\phi\left(  \mathbf{q}%
_{i},\mathbf{q}_{j}\right)  $ and external potential $v\left(  \mathbf{q}%
_{i}\right)  $ is left general at this point. The equilibrium thermodynamics
for this system in the Grand Canonical ensemble is determined from the grand
\ potential $\Omega(\beta\mid\mu,\phi)$
\begin{equation}
\beta\Omega(\beta\mid\mu,\phi)=-\ln\sum_{N}Tr_{N}e^{-\beta\left(  K+\Phi-\int
d\mathbf{r}\mu(\mathbf{r})\widehat{n}(\mathbf{r})\right)  }. \label{2.2}%
\end{equation}
Here the local chemical potential $\mu(\mathbf{r})$ is defined by%
\begin{equation}
\mu(\mathbf{r})=\mu-v(\mathbf{r}), \label{2.3}%
\end{equation}
and the operator $\widehat{n}(\mathbf{r})$ representing the microscopic
density is defined by%
\begin{equation}
\widehat{n}(\mathbf{r})=\sum_{i=1}^{N}\delta\left(  \mathbf{r}-\mathbf{q}%
_{i}\right)  . \label{2.4}%
\end{equation}
The notation $\Omega(\beta\mid\mu,\phi)$ indicates that it is a function of
the inverse temperature $\beta^{-1}=k_{B}T$ and a functional of $\mu
(\mathbf{r})$ and the pair potential $\phi\left(  \mathbf{r},\mathbf{r}%
^{\prime}\right)  $.

A corresponding classical system is defined with a classical grand potential
\begin{equation}
\beta\Omega_{c}(\beta_{c}\mid\mu_{c},\phi_{c})=-\ln\sum_{N}\frac{1}%
{\lambda_{c}^{3N}N!}\int d\mathbf{q}_{1}..d\mathbf{q}_{N}e^{-\beta_{c}\left(
\Phi_{c}-\int dr\mu_{c}(r)\widehat{n}(r)\right)  }, \label{2.5}%
\end{equation}
where $\lambda_{c}=\left(  2\pi\beta_{c}\hbar^{2}/m\right)  ^{1/2}$ is the
thermal de Broglie wavelength and the inverse temperature of the quantum system $\beta$ 
 multiplies the classical grand potential. The classical grand potential is 
defined in terms of an effective inverse
classical temperature $\beta_{c}$, effective classical local chemical
potential $\mu_{c}(\mathbf{r})$, and effective classical pair potential
$\phi_{c}\left(  \mathbf{r},\mathbf{r}^{\prime}\right)  .$ The classical
system therefore has one undetermined scalar and two undetermined functions.
These are defined by the following three conditions%
\begin{equation}
\Omega_{c}(\beta_{c}\mid\mu_{c},\phi_{c})\equiv\Omega(\beta\mid\mu
,\phi),\hspace{0.25in}\frac{\delta\Omega_{c}(\beta_{c}\mid\mu_{c},\phi_{c}%
)}{\delta\mu_{c}(\mathbf{r})}\mid_{\beta_{c},\phi_{c}}\equiv\frac{\delta
\Omega(\beta\mid\mu,\phi)}{\delta\mu(\mathbf{r})}\mid_{\beta}, \label{2.6}%
\end{equation}%
\begin{equation}
\frac{\delta\Omega_{c}(\beta_{c}\mid\mu_{c},\phi_{c})}{\delta\phi
_{c}(\mathbf{r},\mathbf{r}^{\prime})}\mid_{\beta_{c},\mu_{c}}=\frac
{\delta\Omega(\beta\mid\mu,\phi)}{\delta\phi(\mathbf{r},\mathbf{r}^{\prime}%
)}\mid_{\beta,\mu}. \label{2.7}%
\end{equation}
An equivalent form for these conditions can be expressed in terms of the
pressure, the local average density, and the pair correlation function%
\begin{equation}
p_{c}(\beta_{c}\mid\mu_{c},\phi_{c})\equiv p(\beta\mid\mu,\phi),\hspace
{0.25in}n_{c}(\mathbf{r;}\beta_{c}\mid\mu_{c},\phi_{c})\equiv n(\mathbf{r;}%
\beta\mid\mu,\phi), \label{2.8}%
\end{equation}%
\begin{equation}
g_{c}(\mathbf{r},\mathbf{r}^{\prime};\beta_{c}\mid\mu_{c},\phi_{c})\equiv
g(\mathbf{r},\mathbf{r}^{\prime};\beta\mid\mu,\phi). \label{2.9}%
\end{equation}
In this way the classical system has the same thermodynamics and structure as that of
the underlying quantum system.

These definitions for $\beta_{c},\mu_{c},$ and $\phi_{c}$ are only implicit
and require inversion of the classical expressions on the left sides of these
equations to express them in terms of the given quantum variables $\beta,\mu,$
and $\phi.$ Generally this is a difficult classical many-body problem. In
addition, the inversion is expressed in terms of the corresponding quantum
functions $p(\beta\mid\mu,\phi)$, $n(\mathbf{r;}\beta\mid\mu,\phi),$ and
$g(\mathbf{r},\mathbf{r}^{\prime};\beta\mid\mu,\phi)$ which require solution
to the original difficult quantum many-body problem. Hence it would appear
that the introduction of a representative classical system to calculate the
thermodynamics of the quantum system is circular. However, it is expected that
the inversion can be accomplished in some simple approximation that
incorporates relevant quantum effects and the resulting approximate classical
parameters $\beta_{c},\mu_{c},$ and $\phi_{c}$ used in a more accurate theory
or simulation to "bootstrap" a better thermodynamics and structure. This is
illustrated in the next two sections.

\section{Pair Correlations in the Uniform Electron Gas}

\label{sec3}To illustrate the utility and effectiveness the effective
classical system approach defined above, the calculation of pair correlations
in the uniform electron gas is described in this section. The objective is to
describe these correlations over the entire density and temperature plane. In
the classical domain this system is typically known as the one component
plasma. The classical system is well-described by classical methods such as
liquid state theory, classical Monte Carlo, and molecular dynamics simulation.
Both fluid and solid equilibrium phases are now well characterized, including
very strong coupling conditions. Consequently, there is a great potential to
apply these approaches as well to quantum systems using the classical map.

In this section the system of interest is the uniform electron gas at
equilibrium. It is comprised of electrons in a uniform neutralizing background.
The dimensionless temperature used here is the temperature relative to the
Fermi temperature, $t\equiv1/\beta\epsilon_{F}$, where the Fermi energy is
defined by $\epsilon_{F}=\hbar^{2}\left(  3\pi^{2}n\right)  ^{2/3}/2m$ . Also
the density dependence is characterized by the ratio of the average distance
between particles relative to the Bohr radius, $r_{s}\equiv r_{0}/a_{B}$,
where $4\pi nr_{0}^{3}/3=1$ and $a_{B}=\hbar^{2}/me^{2}$. The dimensionless
space scale is $\mathbf{r}^{\ast}\mathbf{=r}/r_{0}$. The classical pair
correlation function at uniform equilibrium depends only on the relative
coordinate so%
\begin{equation}
g_{c}(\mathbf{r},\mathbf{r}^{\prime};\beta_{c}\mid\mu_{c},\phi_{c})\equiv
g_{c}(\left\vert \mathbf{r}^{\ast}-\mathbf{r}^{\ast\prime}\right\vert
,r_{s}\mid\phi_{c}^{\ast}), \label{3.1}%
\end{equation}
where $\phi_{c}^{\ast}\equiv\beta\phi^{\ast}=\Gamma/\left\vert \mathbf{r}%
^{\ast}-\mathbf{r}^{\ast\prime}\right\vert $. Here $\Gamma=\beta e^{2}/r_{0}$
is the classical Coulomb coupling constant. In terms of $t,r_{s}$ it is
$\Gamma=2\left(  \frac{9}{4}\pi\right)  ^{-2/3}r_{s}t^{-1}$. Note that the
functional $g_{c}(r,r_{s}\mid\cdot)$ is independent of $t$. In contrast, the
quantum pair correlation functional depends on both $r_{s}$ and $t$
\begin{equation}
g(\mathbf{r},\mathbf{r}^{\prime};\beta\mid\mu_{c},\phi_{c})\equiv g(\left\vert \mathbf{r}^{\ast}-\mathbf{r}^{\ast\prime}\right\vert,r_{s},t\mid\phi^{\ast}). \label{3.2}%
\end{equation}

Now, using the equivalence of the classical and quantum pair correlation
functions\ (\ref{2.9}) the classical functional can be inverted to give
$\phi_{c}^{\ast}$%
\begin{equation}
\phi_{c}^{\ast}\left(  r^{\ast},r_{s},t\right)  =g_{c}^{-1}(r^{\ast},r_{s}\mid
g). \label{3.3}%
\end{equation}
This is the formally exact definition of the effective classical pair
potential. The practical procedure is to evaluate this in some reasonable,
simple approximation and then "bootstrap" the result in a more sophisticated
approximation to $g_{c}(r^{\ast},r_{s}\mid\phi_{c}^{\ast}).$ First, it is
required that the limit of non-interacting particles be given correctly, so
the potential is written as
\begin{equation}
\phi_{c}^{\ast}\left(  r^{\ast},r_{s},t\right)  =\phi_{c}^{\ast(0)}\left(
r^{\ast},r_{s},t\right)  +\Delta\left(  r^{\ast},r_{s},t\right)  . \label{3.4}%
\end{equation}
Here $\phi_{c}^{\ast(0)}\left(  r^{\ast},r_{s},t\right)  $ is an effective
pair interaction chosen such that its classical pair correlation function is
the same as that for the quantum system with no Coulomb interactions,
$g^{(0)}\left(  r^{\ast},r_{s},t\right)  .$ The second term, $\Delta\left(
r^{\ast},r_{s},t\right)  $, replaces the Coulomb interaction by a
corresponding classical pair interaction incorporating the quantum effects.
Here, it is constrained to be exact in the weak coupling limit. Classically, the
latter corresponds to the potential becoming the same as the direct
correlation function
\begin{equation}
\phi_{c}^{\ast}\left(  r^{\ast},r_{s},t\right)  \rightarrow-c\left(  r^{\ast
},r_{s},t\right)  ,\hspace{0.25in}\phi_{c}^{\ast(0)}\left(  r^{\ast}%
,r_{s},t\right)  \rightarrow-c^{(0)}\left(  r^{\ast},r_{s},t\right)
\label{3.5}%
\end{equation}
where the direct correlation function is defined by the Ornstein-Zernicke
equation%
\begin{equation}
c\left(  r^{\ast},r_{s},t\right)  =g\left(  r^{\ast},r_{s},t\right)
-1-\frac{3}{4\pi}\int d\mathbf{r}^{\prime}c\left(  r^{\ast\prime}%
,r_{s},t\right)  \left(  g\left(  \left\vert \mathbf{r}^{\ast}-\mathbf{r}%
^{\ast\prime}\right\vert ,r_{s},t\right)  -1\right)  . \label{3.6}%
\end{equation}
The quantum weak coupling limit is the random phase approximation. Therefore
(\ref{3.6}) is calculated by inserting the finite temperature random phase
approximation, $g^{RPA}\left(  r^{\ast},r_{s},t\right)  ,$ on the right side.
The resulting approximate effective pair potential is now (\ref{3.4}) with%
\begin{equation}
\phi_{c}^{\ast}\left(  r^{\ast},r_{s},t\right)  \simeq\phi_{c}^{\ast
(0)}\left(  r^{\ast},r_{s},t\right)  -c^{RPA}\left(  r^{\ast},r_{s},t\right)
+c^{(0)}\left(  r^{\ast},r_{s},t\right)  . \label{3.7}%
\end{equation}
Clearly this incorporates the ideal gas and weak coupling limits, without
being restricted to either.

The qualitative differences of this effective classical potential from the
underlying Coulomb potential of the quantum system are two fold. First, the
divergence at $r^{\ast}=0$ is removed, i.e. $\phi_{c}^{\ast}\left(  r^{\ast
}=0,r_{s},t\right)  $ is finite. Second, for large $r^{\ast}$ the potential is
also of the Coulomb form, but with a different amplitude
\begin{equation}
\phi_{c}^{\ast}\left(  r^{\ast},r_{s},t\right)  \rightarrow\Gamma_{e}\left(
t,r_{s}\right)  r^{\ast-1}, \label{6}%
\end{equation}
The classical Coulomb coupling constant $\Gamma\left(  t,r_{s}\right)  $ \ has
been replaced by the effective quantum coupling constant $\Gamma_{c}\left(
t,r_{s}\right)  $%
\begin{equation}
\Gamma_{e}\left(  t,r_{s}\right)  =\frac{2}{\beta\hbar\omega_{p}\coth\left(
\beta\hbar\omega_{p}/2\right)  }\Gamma,\hspace{0.25in}\Gamma\left(
t,r_{s}\right)  =\beta q^{2}/r_{0}=2\left(  \frac{9\pi}{4}\right)^{-2/3}\frac{r_{s}}{t}
\label{7}%
\end{equation}
Here $\beta\hbar\omega_{p}=\beta\hbar\left(  4\pi nq^{2}/m\right)  ^{1/2}$
=$4\left(  2\sqrt{3}\pi^{-2}\right)  ^{1/3}r_{s}^{1/2}/3t$ is the
dimensionless plasma frequency.

With the pair potential determined in this way the pair correlation function
$g\left(  r^{\ast},r_{s},t\right)  $ can be calculated beyond the ideal gas and
weak coupling conditions using, for example, molecular dynamics or classical
Monte Carlo simulation. Here the results are illustrated using an integral
equation from liquid state theory or classical density functional theory. It
is the hypernetted chain approximation (HNC)%
\begin{equation}
\ln g\left(  r^{\ast},r_{s},t\right)  =-\phi_{c}^{\ast}\left(  r^{\ast}%
,r_{s},t\right)  -c\left(  r^{\ast},r_{s},t\right)  +\left(  g\left(  r^{\ast
},r_{s},t\right)  -1\right)  \label{3.8}%
\end{equation}

This equation together with the Ornstein-Zernicke equation (\ref{3.6})
provides a coupled set of equations for both $c\left(  r^{\ast},r_{s}%
,t\right)  $ and $g\left(  r^{\ast},r_{s},t\right)  .$ Figures 1 and 2 show
the results in comparison with recent path integral Monte Carlo (PIMC) simulations
\cite{Brown} at $r_{s}=1$ and $6$ for a wide range of $t$. Clearly there is
quite good agreement with this benchmark data using this standard liquid state
classical theory modified only by the quantum effects in the modified pair
potential. For additional details and other values for $r_{s},t$ see reference
\cite{DD13}.

\begin{figure}[htb]
\begin{minipage}[t]{80mm}
\includegraphics[width=\columnwidth]{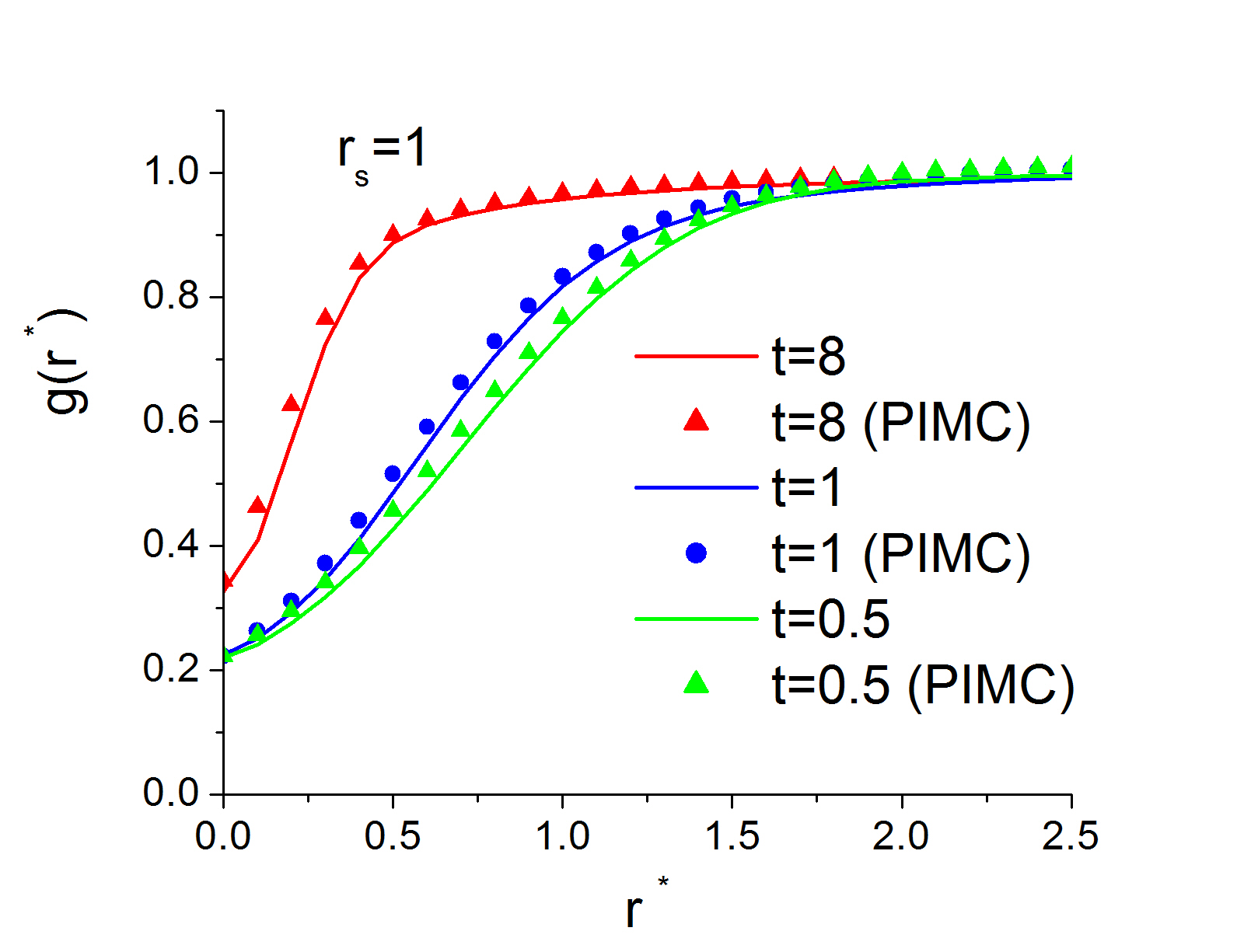}%
\caption{Pair correlation function $g(r^*)$ for $r_s=1$ at $t=$ 0.5, 1, and 8. }
\label{fig1}
\end{minipage}
\hspace{\fill}
\begin{minipage}[t]{75mm}
\includegraphics[width=\columnwidth]{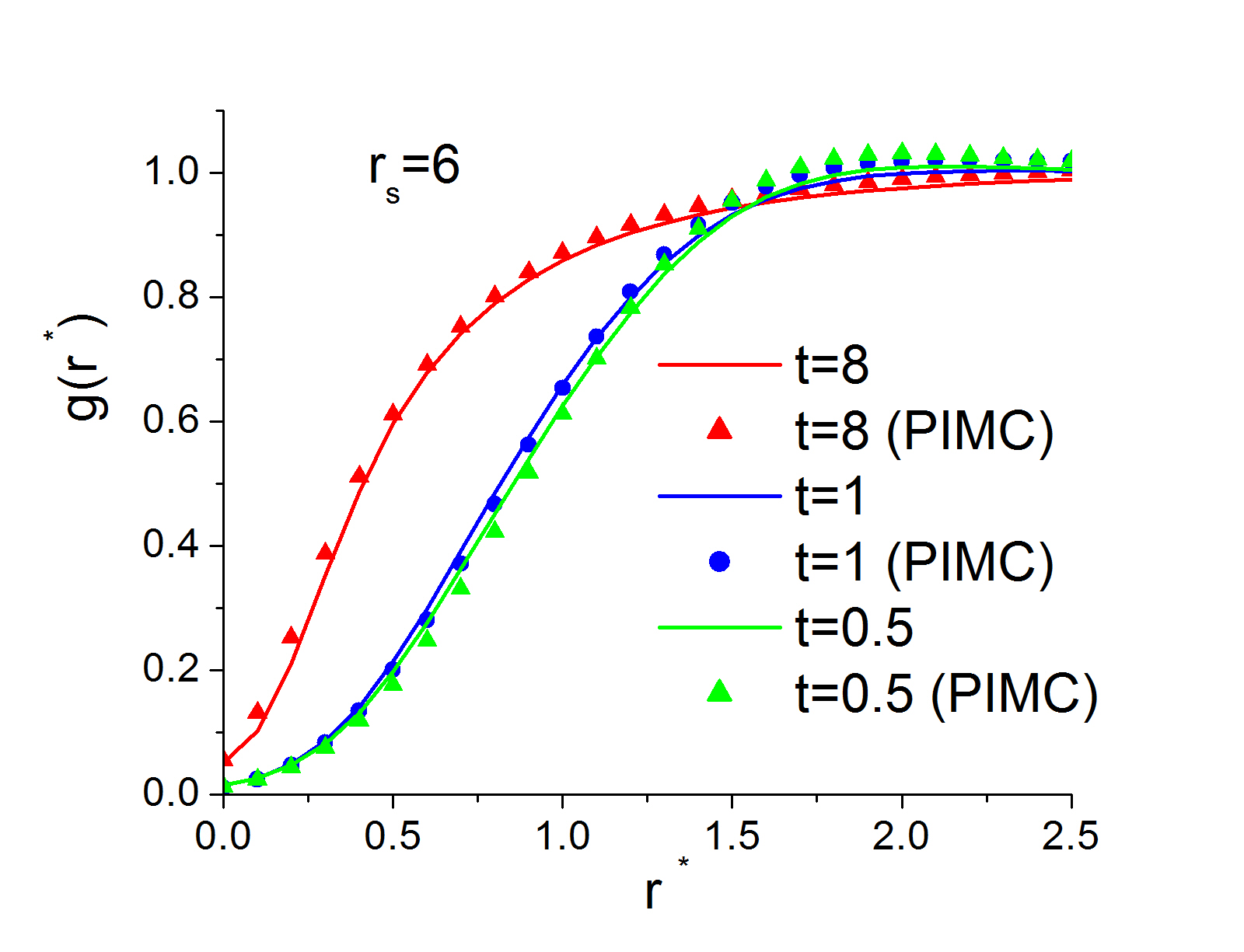}
\caption{Pair correlation function $g(r^*)$ for $r_s=6$ at $t=$ 0.5, 1, and 8. }
\label{fig2}
\end{minipage}
\end{figure}

\section{Charges in a Harmonic Trap}

\label{sec4}For a second application of the classical map consider $N$
charges in a harmonic trap. The classical HNC of the last section, extended to
this inhomogeneous system \cite{Attard89}, has been shown to give an accurate
description of the radial density profile for this system \cite{Wrighton}. Of
particular interest is the quantitative description of shell structure that
occurs for classical strong coupling conditions. In this section, that approach is extended to include quantum effects.

The HNC theory for the density profile together with (\ref{2.8}) and (\ref{2.9}) leads to%

\begin{equation}
\ln\left(  n\left(  \mathbf{r}\right)  \lambda_{c}^{3}\right)  =\beta_{c}%
\mu_{c}-\beta v_{c}(\mathbf{r})+\int d\mathbf{r}^{\prime}c(\left\vert
\mathbf{r-r}^{\prime}\right\vert ,\mu_{c},\beta_{c})n\left(  \mathbf{r}%
^{\prime}\right)  . \label{4.1}%
\end{equation}
Here $c(r,\mu_{c},\beta_{c})$ is the same direct correlation function as
described in the previous section. Also, $v_{c}(\mathbf{r})$ is the effective
classical trap which includes quantum behavior of the given harmonic trap. The
approach is to determine $v_{c}(\mathbf{r})$ approximately by inverting
(\ref{4.1}) for a given approximate quantum density $n\left(  \mathbf{r}%
\right)  $. Two approximations are compared here.

The first approximation is to invert (\ref{4.1}) for non-interacting particles
in a trap. The quantum effects in this case are entirely due to exchange
symmetry%
\begin{equation}
v_{c}^{\ast(0)}(\mathbf{r}^{\ast})=c-\ln\left(  n^{\ast(0)}\left(
\mathbf{r}^{\ast}\right)  \right)  +\int d\mathbf{r}^{\ast\prime}%
c^{(0)}(\left\vert \mathbf{r}^{\ast}\mathbf{-r}^{\ast\prime}\right\vert
,r_{s},t)n^{\ast(0)}\left(  \mathbf{r}^{\ast\prime}\right)  . \label{4.2}%
\end{equation}
Here $v_{c}^{\ast}(\mathbf{r}^{\ast})\equiv$ $\beta_{c}v_{c}(\mathbf{r})$ and
the superscript $0$ on a property denotes its ideal gas value. Also, $c$ is a
constant that only sets the normalization of the density profile. The
calculation of $n^{\ast(0)}\left(  \mathbf{r}^{\ast}\right)  $ is
straightforward in terms of the harmonic oscillator eigenfunctions, but for
the case considered here $\left(  N=100\right)  $ it is found that the local
density approximation (finite temperature Thomas-Fermi) is quite accurate. An
important qualitative feature of $n^{\ast(0)}\left(  \mathbf{r}^{\ast}\right)
$ is its vanishing at a finite $\mathbf{r}^{\ast}$ as $t\rightarrow0$. This
leads to the formation of a hard wall in the effective classical trap. It is
well known that such hard walls produce shell structure in classical
mechanics, so this represents a quantum origin for new shell structure
independent of Coulomb correlations.

The second approximation is to invert (\ref{4.1}) with mean field quantum
Coulomb correlations%
\begin{equation}
v_{c}^{\ast(H)}(\mathbf{r}^{\ast})=c-\ln\left(  n^{\ast(H)}\left(
\mathbf{r}^{\ast}\right)  \right)  +\int d\mathbf{r}^{\ast\prime}\left(
c^{(0)}(\left\vert \mathbf{r}^{\ast}\mathbf{-r}^{\ast\prime}\right\vert
,r_{s},t)-\frac{\Gamma\left(  t,r_{s}\right)  }{\left\vert \mathbf{r}%
^{\ast}\mathbf{-r}^{\ast\prime}\right\vert }\right)  n^{\ast(H)}\left(
\mathbf{r}^{\ast\prime}\right)  . \label{4.3}%
\end{equation}
The density $n^{\ast(H)}\left(  \mathbf{r}^{\ast}\right)  $ is calculated from
quantum density functional theory without exchange or correlation (Hartree
approximation), and again using the local density approximation. It gives a
qualitative change from the ideal gas form (\ref{4.2}) since the system is
considerably expanded by the Coulomb repulsion. The hard wall is mitigated and
resulting effective classical trap potential has a more harmonic form.

Figure 3 shows the density profiles obtained from (\ref{4.1}) using
(\ref{4.2}) or (\ref{4.3}). It is seen that the strong shell structure from
the ideal gas hard wall is removed when Coulomb interactions are included.
Also shown on this figure are the effective classical trap potentials for the
two cases. The result from (\ref{4.2}) shows a kink which is a precursor of
the hard wall at $t=0$. Although this deviation from harmonic is small, it is
sufficient to generate a large shell. In contrast, the result from (\ref{4.3})
is more nearly harmonic and has only the shell due to Coulomb correlations
already present in the purely classical calculation (no quantum effects) \cite{Wrighton}. 
The lesson from this comparison is that quantum effects on the
effective classical trap potential determined without Coulomb interactions
lead to a false mechanism for shell structure. The more realistic mean field
quantum determination does not have this shell structure and provides a quite
different density profile. A more complete discussion of this comparison and
results for a wide range of $r_{s},t$ will be given elsewhere \cite{WDD15}.

\begin{figure}[htb]
\begin{minipage}[t]{80mm}
\includegraphics[width=\columnwidth]{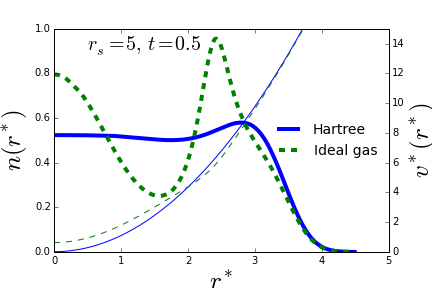}
\caption{Density profiles (thick lines) and effective classical trap potentials (thin lines) using the Hartree and ideal gas approximations.}
\label{fig3}
\end{minipage}
\hspace{\fill}
\end{figure}

\section{Discussion}

\label{sec5}The use of an effective classical system to describe quantum
effects has been shown to provide a practical tool by the two examples of the
previous sections. Although the results are quite good for the methods used
here to determine the effective pair potential and trap potential, and the HNC
implementation of the classical statistical mechanics, improvements in both
remain to be explored. For example, the limitations of the HNC theory can be
eliminated using these same potentials in molecular dynamics or Monte Carlo
simulations. 

The comparisons of the last section show that incorrect results
can be obtained if the input quantum mechanics for the effective potentials is
not sufficiently representative of the real system. For the electron gas, the
effective potential has a Coulomb tail whose amplitude is constrained to
satisfy an exact sum rule. It would be useful to have exact
constraints for other systems as well, to assure applicability over a wide
range of the parameter space.

As noted in the introduction, there is strong current interest in systems of
electrons and ions to describe conditions of warm, dense matter \cite{WDM}.
Such systems are described by molecular dynamics simulation of the ions whose
forces are calculated from a density functional theory (DFT) for the electrons
at each time step. Traditionally, the DFT calculation is performed within the
Kohn-Sham approach requiring a self-consistent diagonalization of an effective
single electron Hamiltonian to construct the local density. At temperatures
approaching the Fermi temperature, the number of relevant states (orbitals)
becomes large and the calculations are no longer practical. A resolution of
this problem is to forgo the Kohn-Sham method and return to the original form
of DFT with a single Euler equation for the local density determined from the
free energy as a known functional of the density. The primary difficulty is
finding the non-interacting free energy as a functional of the density, which
remains an unsolved problem in the quantum theory. However, its classical
counterpart does not have this difficulty -- the non-interacting free energy
is known as an explicit functional of the density. Hence, an implementation of
the effective classical system as described here, together with classical DFT,
provides the desired orbital free DFT.

To see how this might be implemented consider a system of $N_{e}$ electrons
and $N_{i}$ positive ions with charges $Z$ and positions
$\left\{  \mathbf{R}_{\alpha}\right\}  $. For charge neutrality $N_{i
}Z=N_{e}$. This can be viewed as an electron system in the external
potential of the ions
\begin{equation}
v(\mathbf{r})=\sum_{\alpha=1}^{N_{i}}\frac{-Ze^{2}}{\left\vert
\mathbf{r}-\mathbf{R}_{\alpha}\right\vert }\;. \label{5.1}%
\end{equation}
Return to (\ref{4.1}) for the corresponding local electron density, where now
$v_{c}(\mathbf{r})$ is the effective classical potential corresponding to
(\ref{5.1}). A reasonable, realistic determination of $v_{c}(\mathbf{r})$
might be given by (\ref{4.3}) in the form%
\begin{align}
v_{c}^{\ast(HF)}(\mathbf{r}^{\ast})&=c-\ln\left(  n^{\ast(HF)}\left(
\mathbf{r}^{\ast}\right)  \right)\nonumber\\
&\qquad  +\int d\mathbf{r}^{\ast\prime}\left(
c^{(0)}(\left\vert \mathbf{r}^{\ast}\mathbf{-r}^{\ast\prime}\right\vert
,r_{s},t)-\frac{\Gamma\left(  t,r_{s}\right)  }{\left\vert \mathbf{r}%
^{\ast}\mathbf{-r}^{\ast\prime}\right\vert }\right)  n^{\ast(HF)}\left(
\mathbf{r}^{\ast\prime}\right)  , \label{5.2}%
\end{align}
where now $n^{\ast(HF)}\left(  \mathbf{r}^{\ast}\right)  $ is the Hartree-Fock
electron density for the given array of electrons. While determination of
$n^{\ast(HF)}\left(  \mathbf{r}^{\ast}\right)  $ is still non-trivial it is a
practical problem, and then use of $v_{c}^{\ast(HF)}(\mathbf{r}^{\ast})$ in
(\ref{4.1}) gives the desired orbital free DFT for the electrons.

Dharma-wardana has proposed a more complete application of the classical DFT
for both the electrons and ions \cite{DW}, eliminating the molecular dynamics
simulation for the ions. An additional effective classical electron - ion
potential must be determined in this case.

\section{Acknowledgments}

The authors are indebted to Michael Bonitz for his comments and criticism of
an earlier ms. This research has been supported in part by NSF/DOE Partnership
in Basic Plasma Science and Engineering award DE-FG02-07ER54946 and by US DOE
Grant DE-SC0002139.

\end{document}